\documentclass[10pt,conference]{IEEEtran}
\usepackage{mwe}
\usepackage{hyperref}
\usepackage{multirow}
\usepackage[numbers]{natbib}
\begin{document}

\title{JavaBERT: Training a transformer-based model for the Java programming language}

\author{\IEEEauthorblockN{Nelson Tavares de Sousa, Wilhelm Hasselbring}
	\IEEEauthorblockA{\textit{Software Engineering Group, Kiel University}\\
		Kiel, Germany \\
		\{tavaresdesousa, hasselbring\}@email.uni-kiel.de}
}

\maketitle

\begin{abstract}
Code quality is and will be a crucial factor while developing new software code, requiring appropriate tools to ensure functional and reliable code.
Machine learning techniques are still rarely used for software engineering tools,
missing out the potential benefits of its application.
Natural language processing has shown the potential to process text data regarding a variety of tasks.
We argue, that such models can also show similar benefits for software code processing.
In this paper, we investigate how models used for natural language processing can be trained upon software code.
We introduce a data retrieval pipeline for software code and train a model upon Java software code.
The resulting model, JavaBERT, shows a high accuracy on the masked language modeling task showing its potential for software engineering tools.
\end{abstract}

\section{Introduction}
Reliability is gaining importance as more and more systems across all domains, such as Internet of Things devices and banking systems, rely on software to function properly.
Defective software has the potential to cause malfunctioning systems which may impose a variety of risks and greater amount of costs.
Therefore, software quality is a significant factor which must be considered during software development, playing a crucial role in software engineering (SE).
Rule-based systems, like static code analysis, allow to detect bugs, code smells, and similar issues, improving code quality \cite{Mueller2016}. 
However, they rarely are able to self-adapt to evolving environments, as this requires a manual update of the rules.
Machine learning (ML) could close this gap by learning from such new environments, but we still observe a lack of usage of ML technologies in the domain of SE.
SE just starts to make use of AI powered tools, but still stays behind the advancements made in other ML domains such as natural language processing (NLP).

NLP has seen drastic improvements over the past years with the development of new ML models, especially with transformer-based models.
Such models show a high performance, which is able to outperform a human baseline on certain tasks \cite{Rajpurkar2016,Wang2018}.
Therefore, the state-of-the-art of NLP already proves to be reliable on performing tasks upon the semantic content of natural language.
First approaches to take concepts from the NLP domain to the domain of program code understanding emerge \cite{Feng2020}.
Yet current approaches handle software code as raw text neglecting the syntactical features given by programming languages.
Additionally, most existing approaches are tailored to a specific downstream task, narrowing down their field of application.
In contrast to pre-trained models, which are trained on generic tasks, potential further research with such specific models is limited.

We argue that not only a well pre-trained model for software code is required, but also a defined processing pipeline to improve reproducibility.
Thanks to the application of transfer learning, such models carry the potential to accelerate the development of new approaches.
Different downstream tasks can be trained on the same baseline model \cite{BERT}, increasing the performance and comparability of the resulting models.
An adaptable pipeline can also be used for different programming languages, facilitating the incorporation of new languages. 

In this paper, we present our approach to train transformer-based models on source code to enable the use of ML for software engineering.
We contribute by proposing a data retrieval pipeline for software code and a concept which allows to take syntactical features explicitly into account while training a model 
Additionally, we evaluate whether this concept improves the final performance.
Lastly, we present a state-of-the-art model for the Java programming language, called JavaBERT.

\section{Related Work}
Recent research in this domain shows an increasing interest in the application of machine learning for programming tasks.

For instance, the approach presented by \citet{Tufano_2021} gives an example on how ML can be used in the domain of SE.
To assist the development process, an ML model is used to anticipate recommended changes in reviews.
With a multimodal approach, program code before and after those changes, and review comments are used to train the modal.

\citet{bielik2017learning} show how machine learning can be used for static code analysis.
Here, a ML model tries to identify patterns which are associated with malicious program behavior or code smells
This is achieved by introducing a language-agnostic rule specification upon which the model is trained to identify code parts which violate these rules.
While being able to be applied to different languages, the rule set still needs to be updated manually if changes regarding the detectable code parts need to be included.

By combining a convolutional neural network (CNN) with a long short-term memory (LSTM), \citet{Huo2017} were able to achieve state-of-the-art performance for bug detection.
Their contribution especially relies on the application of a LSTM to exploit the sequential nature of program code.
This approach is then fed with bug reports such that the model is able to be trained toward bug detection.
In contrast to this approach, we will employ models which are able to grasp context to a greater extent, eliminating the requirement for a LSTM.
The combination of a CNN and LSTM by \citet{Huo2017} is tailored to the use case of bug detection, whereas we aim to pretrain a model on a generic task making it suitable for downstream tasks later on.

The Bidirectional Encoder Representations from Transformers (BERT) model is able to capture the context of large text documents and propagate the context throughout the whole document to both sides \cite{BERT}.
Therefore, any token in the document is able to attend the context to its left and right side.
This allows BERT to be trained on documents of greater sizes with comparably small models.
The architectures of previous state-of-the-art models imposed a lower overall limit to the document sizes.
Furthermore, BERT also uses masked language modeling (MLM) for its pretraining task.
With MLM, words in the input are masked and the model's task is to predict the original word.
BERT demonstrated how this task is able to pretrain a reliable model which can be used to train for downstream tasks and show state-of-the-art performance.
We transfer BERT to Java code.

\citet{Feng2020} used data sets to first pretrain a BERT model relying on a variety of programming languages and afterwards trained it upon several downstream tasks.
The resulting model shows state-of-the-art performance on the given tasks, showing how transformer-based models like BERT may also be applicable to other tasks in the domain of software development.
In contrast to this work, we investigate how a model dedicated to a specific language performs compared to such a broader approach.

\section{Research Objectives}\label{sec:ro}
Our goal is to research ways to pretrain a BERT-based model for programming languages, in our case Java.
Three research questions (RQ) can be derived to give an answer to this.

\subsection*{RQ1: How can program code be retrieved for training purposes?}
A proper amount of data is crucial to train any ML model.
Large amounts of corpora are made available for the domain of NLP but such data sets still lack for programming languages.
Sources for a greater amount of programming code files are available but still need to be collected and be bundled into corpora.
We will showcase how this can be done for the Java language and collect data for our use case.

\subsection*{RQ2: How to train an NLP model for programming languages?}
Data for training purposes undergo certain preprocessing steps before being forwarded to an ML model.
However, in our use case, program code may benefit from an adapted approach.
Syntactical features of programming languages may be used to increase the efficiency of ML models for such domains.
We will introduce such an approach, evaluate it by comparing the results with state-of-the-art approaches, and discuss the advantages and disadvantages of it.
\subsection*{RQ3: How does such a programming language model perform?}
To evaluate the merits of such an approach, we need to measure the performance of our approach.
Metrics will be introduced which give a well-defined measurement that can be used to compare models.
These metrics will be applied to all trained models to evaluate and compare the performance of our custom approach with state-of-the-art NLP models.

\section{Architecture}
\subsection{Data Retrieval}
\begin{figure*}[htb]
	\centering
	\includegraphics[width=\textwidth]{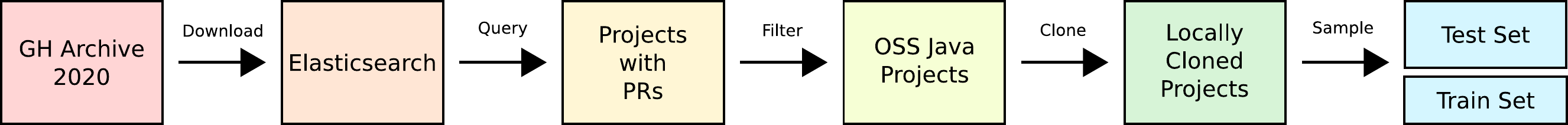}
	\caption{Data retrieval and filtering pipeline}
	\label{fig:retrieval}
\end{figure*}
The first step consists of the selection of appropriate data for our use case.
First of all, we choose a specific language we want to use.
The sole requirement for this language is its broad acceptance and usage.
The PYPL ranking\footnote{\url{https://pypl.github.io/PYPL.html}} gives an insight into the programming languages which have been most searched on Google.
For the year 2021, Python takes first place, however Java has constantly shown a higher ranking until recent years.
Similar information can be taken from the TIOBE index.\footnote{\url{https://www.tiobe.com/tiobe-index/}}
We chose Java as our target language, due to its high rankings in the previous years, which will most likely yield in greater amounts of available code in comparison to Python or other languages.

As a next step, we decide upon a data source.
We require lots of publicly available data which can be reused for our purposes.
GitHub, as a platform providing a version control service, provides a multitude of Open Source Software (OSS), which complies to these requirements.
An addition to that, GitHub provides interfaces (APIs), which allow to investigate each repository, without downloading all data and parsing it beforehand.
Therefore, we chose GitHub as our data source.

After this preliminary decision, we start to retrieve the data.
\autoref{fig:retrieval} gives an overview of the different steps of our retrieval pipeline.
Each square depicts the status of the data throughout the pipeline.
The different processing and filtering actions are depicted by arrows.
The GH Archive project\footnote{\url{https://www.gharchive.org/}} allows to download all API events which took place in GitHub for specified time ranges.
We download all data for the year 2020 and load it into an Elasticsearch index, which allows us to perform various queries on the data without relying on third party services.

We choose to filter for projects which show recent activities.
Pull requests give a simple metric for active projects.
Therefore, we will first only take projects into account which have yielded pull requests.
We do so, by performing an aggregation query on our Elasticsearch instance, which counts the events triggered by the creation of comments within pull requests for each project.
Projects without any pull request will not be returned by this query.
This list of active projects is further filtered down by their used licenses and programming languages.
To do so, we use the provided API by GitHub, which returns metadata regarding the project, such as the used license and programming languages.
We filter out projects which do not make use of the MIT or Apache 2.0 license and in which Java is not within top three used programming languages.
Choosing these licenses allows to avoid conflicts regarding copyright issues and improves the availability of the data for further research. 
In addition, we filter out all projects, which have less than 10 comments within all pull requests.
Eventually, this yields in 26983 projects which are then cloned to a local drive.

In a next step, we take a sample out of all Java files.
This includes a simple filtering of smaller and bigger files.
As threshold we decide to filter out files with more than 3000 and less than 40 Java tokens.
This eliminates files with almost no content, such as empty classes, and also god classes with greater amount of code content.
As a result, we have a sample of the size of 4283372 Java files.
These are split into two sets which represent the training and test sets.
The training set contains a random subset of around 70\% (2,998,345) of all Java files, whereas the test set contains the remaining 1,285,027 files of those files.

\subsection{Tokenizer}
Tokenization describes a preprocessing step, where text is divided into tokens which then can be fed into ML models.
In NLP this means typically the separation by words or smaller parts of text like subword units.
The latter allows to deal with the disadvantages of fixed vocabularies \cite{Wu2016}.
Afterwards, tokens are assigned a unique numeric identifier such that they can be processed by ML models.

Programming languages undergo a different segmentation by using lexemes, which introduce semantics to sequences of characters \cite{Aho2007}.
These lexemes are then analyzed and further processed into a stream of tokens.
For Java, the Java Language Specification (JLS) defines how such tokens may be composed.\footnote{\url{https://docs.oracle.com/javase/specs/jls/se8/html/jls-3.html\#jls-3.5}}.
Tokens can be further distinguished into five types: Identifier, Literal, Keyword, Separator, and Operator.
For our use case, we will divide these into two groups.
Tokens of the types Separator and Operator are used by Java to identify structural or operational entities, such as \emph{\{}, \emph{\}}, \emph{==}, etc..
Keyword tokens include a specific set of tokens, which are reserved by Java itself, such as \emph{if}, \emph{new}, etc...
These token types can be considered fixed, as their textual representations always stay the same.
For our approach, we add the literals \emph{true}, \emph{false}, and \emph{null} to this list, as these seem like keywords, but technically are not considered as such.
We further refer to this token set as special tokens.

Tokens of the types Identifier and Literal compose a second group of non-special tokens.
These are not limited to a specific vocabulary and may reflect any combination of character sequences.
The challenge here is to provide a vocabulary which can be used to tokenize these words.
The number of indexed words is a relevant factor as it imposes some trade-offs.
We may provide tokens for any provided word, however the size of the vocabulary dominates the performance of the tokenizer.
This can be circumvented by further splitting words into smaller segments, effectively reducing the required vocabulary size.
WordPiece \cite{Wu2016} is an algorithms which supports such splitting into smaller segments.
It searches for common segments within all words and splits any words in the text by such segments.
This effectively reduces the vocabulary size, as such subword segments can be reused to compose complete words instead of including all possible segment combinations occurring in the text.
This is especially the case for programming languages, due to the common usage of variable names such as \emph{isEnabled}, \emph{isDisabled}, etc..

\begin{figure}[htb]
	\centering
	\includegraphics[width=.45\columnwidth]{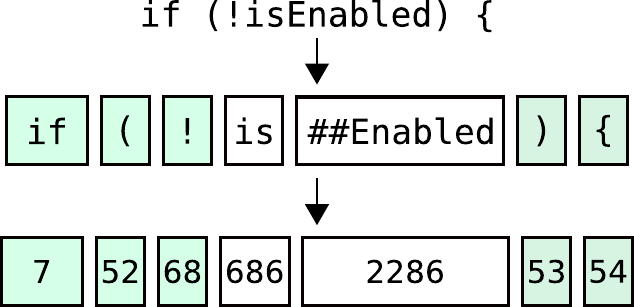}
	\caption{Output of the trained tokenizer with a vocabulary size of 8000}
	\label{fig:tokenizer}
\end{figure}

All files of our sample are used to first extract included tokens.
The \emph{Javalang} python library\footnote{\url{https://github.com/c2nes/javalang}} allows us to do so respecting the rules given by the JLS.
Special tokens are extracted from the JLS and saved in a separate list.
To this list, we add special tokens required by BERT, such as \emph{[MASK]}, \emph{[UNK]} or \emph{[PAD]}.
These tokens are used by the BERT model to indicate masked tokens or tokens uncovered by the provided vocabulary.
This allows WordPiece to handle these tokens separately and not to further split them into subwords.
To train the tokenizer, we use the Tokenize library provided by Hugging Face \cite{huggingface}.
To analyze the performance impact of different vocabulary sizes, we train the tokenizer thrice with different settings for the resulting vocabulary size.
We chose to train the tokenizer on sizes of 8000, 16000, and 32000, which reflect the range most used in the NLP domain.
Furthermore, we want to analyze how the vocabulary sizes will impact the performance of the trained models, as it was already shown to have a greater influence on this aspect \cite{Sennrich_Haddow_Birch_2016}.
\autoref{fig:tokenizer} shows a short example on how the trained tokenizer handles code.
In a pre-tokenization step, the string is separated into lexemes, which then are partially further split into subwords.
Eventually, each token is encoded with a unique id.
The special tokens are highlighted in this example and remain untouched.
\subsection{Model}
We choose BERT as model to train, as it is well-known and also used for other research and optimizations \cite{Lan_Chen_Goodman_Gimpel_Sharma_Soricut_2020,Liu_Ott_Goyal_Du_Joshi_Chen_Levy_Lewis_Zettlemoyer_Stoyanov_2019}.
Similar optimization approaches may be applicable on our final model.
As we perform a MLM training task, we use an untrained BERT model with a language modeling (LM) head.
The model will be trained on program code on which a certain amount of tokens are masked.
Its task is to predict the original tokens.

\section{Experiment}
Our experiment is set up by a training process and a subsequent evaluation.
We will train three different instances of our BERT model, each using one of the three vocabularies.
To examine the possible benefit of the Java tokenizer, we train further BERT instances using the pretrained vocabulary for \emph{BERT Base Uncased} and \emph{BERT Base Cased} to compare our approach with generic ones.
Eventually, an evaluation step will compare the performance of all models against each other.
As we aim to evaluate the performance of different vocabulary sizes, all remaining parameters will remain untouched.
We use the \emph{Transformers} (4.9.1), \emph{Tokenizers} (0.10.3), and \emph{Datasets} (1.10.2) libraries by Hugging Face in combination with \emph{PyTorch} (1.7.1).
Training is performed on three Nvidia Titan X GPUs with 12GB of memory each, running on CUDA 10.1.
Two Intel Xeon E5-2650 with a total of 128GB of memory are used to run the WordPiece algorithm to create the vocabularies.
The configuration for our BERT model relies mainly on the default values given by the \emph{Transformers} library (12-layer, 768-hidden, 12-heads).
The vocabulary size will be set to one of the values of either 8000 (BERT Java-8k), 16000 (BERT Java-16k), 32000 (BERT Java-32k), or 30522 (BERT Base Cased/Uncased) depending on which model we aim to train.
For the optimizer we rely on the default, which is the AdamW optimizer \cite{Loshchilov_Hutter_2019} and use a learning rate of $5e-5$.
A batch size of effectively 30 and a single epoch is used.
With these parameters, training takes between 17 and 24 hours depending on the vocabulary size, with smaller vocabularies resulting in faster training processes.
The loss is calculated by using a cross-entropy criterion.
\begin{figure}[htb]
	\centering
	\includegraphics[width=\columnwidth]{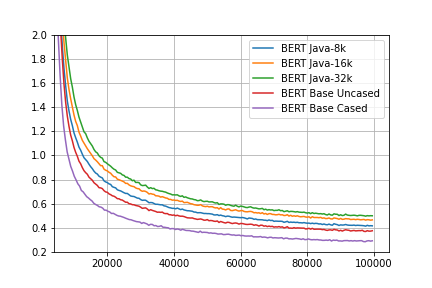}
	\caption{Excerpt of the calculated loss throughout the training process.}
	\label{fig:losses}
\end{figure}
\autoref{fig:losses} shows an excerpt of the loss progress throughout the training for all models.
We omitted the first 4000 steps for the sake of a better differentiation of the losses at later steps in our visualization.
All models perform similar in the first 4000 steps, starting with losses between 5.6 and 5.8.
However, all models converge at slightly different levels.
Models trained with a Java specific tokenizer converge at levels of 0.42 for BERT Java-8k, 0.46 for BERT Java-16k, and 0.5 for BERT Java-32k.
The BERT Base model with an uncased tokenizer shows a better training outcome and converges at 0.36.
With a final loss at 0.29, the BERT Base model with a case-sensitive tokenizer shows the most promising results of all models.

\section{Evaluation}
For the evaluation of all models, we use the test split of our data set, as benchmarking data sets for MLM on Java are not available yet.
To measure the prediction accuracy of the models, we employ the Word Error Rate (WER) metric, but adapt is to the MLM task.
WER is mostly used on large text parts such as in translation tasks and measures the rate of correctly predicted words \cite{morris2004and}.
As only a certain amount of tokens are masked in MLM, the remaining unmasked words can be ignored in our calculation.
Including these words in the metric would always imply perfect matches and would therefore skew the results.
To circumvent this, we simplify the WER calculation and only consider the masked words with:
\[R = \frac{C}{N}\]
The value $N$ depicts the number of masked tokens and $C$ is the number of correctly predicted tokens.
$R$ is the rate of correctly predicted tokens.
For $C$ we follow two approaches to count the correct number of tokens.
With \emph{1-Word-Match} we consider a prediction correct, if the prediction with the highest score matches the original token.
The \emph{3-Words-Match} considers a prediction correct, if one of the first three predictions sorted by highest scores matches the original token.
We ran the test data set on all models and aggregated the results by calculating the mean values over all batch results.
Due to the similar architecture and MLM training task, we included CodeBERT in the evaluation.
Besides its architecture, CodeBERT differs from our approaches, as it is mainly trained upon multiple programming languages with a smaller set of Java files compared to our data set \cite{Feng2020}.

\renewcommand{\arraystretch}{1.2}
\begin{table}[h!]
	\centering
	\begin{tabular}{c|cc} 
		Model & 1-Word-Match & 3-Words-Match\\
		\hline
		BERT Java-8k & 91.8\% & 95.0\% \\ 
		BERT Java-16k & 91.1\% & 94.3\% \\ 
		BERT Java-32k & 90.9\% & 94.0\% \\ 
		BERT Base Uncased & 92.9\% & 95.7\% \\
		BERT Base Cased & \textbf{94.4\%} & \textbf{96.6\%} \\
		CodeBERT & 75.2\% & 81.6\% \\ 
	\end{tabular}
	\bigskip
	\caption{\label{tab:scores}Scores for each ML model}
	\vspace{-1em}
\end{table}

\autoref{tab:scores} shows the results as percentage values, showing the accuracy of the predictions.
Regarding our approach, we see that the \emph{BERT Java-8k}, trained on the smallest vocabulary outperforms the other two Java-specific models.
This meets the observations made on the loss values while training.
\emph{BERT Java-8k} is able to correctly predict 91.8\% of the masked tokens on first try, whereas both bigger models perform slightly worse with 91.1\% and 90.9\%.
The same observation can be made on the \emph{3-Words-Match} score.
However, training BERT with a vocabulary intended for natural language outperforms the models with a Java-specific vocabulary on both scores with up to 94.4\% on the \emph{1-Word-Match}.
A closer inspection shows that WordPiece differentiates syntactical features in the same way as our approach.
However, its vocabulary contains several variants of these tokens which lead to the assumptions that certain sequences may return these non-exclusive variants.
As we were not able to observe this, we argue that the increased performance is mainly achieved by a more efficient tokenization of the literals and identifiers.
We also argue, that the naming convention within Java accounts for the increased performance of the \emph{BERT Base Cased} variant in comparison with \emph{BERT Base Uncased}, which allows the model to better distinguish between different types of entities.
CodeBERT shows a significantly lower scoring performance in comparison to all other models.
This may be explained due to amount of files upon CodeBERT was trained, as well due to its training task upon multiple languages.
As a result, CodeBERT is not able to predict tokens in Java as reliable as dedicated models.

\begin{table}[h!]
	\centering
	\begin{tabular}{c|c}
		Model & \# Parameters \\
		\hline
		BERT Java-8k & 92.19 M \\ 
		BERT Java-16k & 98.35 M \\
		BERT Java-32k & 110.65 M \\
		BERT Base Uncased & 109.51 M \\
		BERT Base Cased & 109.51 M \\
		CodeBERT & 124.70 M \\ 
	\end{tabular}
	\bigskip
	\caption{\label{tab:params}Number of parameters}
	\vspace{-0.5em}
\end{table}
We observed different memory consumption and execution times between the models.
\autoref{tab:params} shows the number of parameters for each model.
The increased memory consumption and execution times correlate with the parameter sizes, therefore underlining the assumption of the model size being the main contributor for this.
However, due to the increased memory footprint, we needed to change the test batch sizes depending on the model, which may also show impact on the execution times.
These aspects even render the bigger models using the Java-specific tokenizer more obsolete, as the increased execution times and memory consumption do not come with greater performance.
Comparing \emph{BERT Java-8k} with \emph{BERT Base Uncased} shows a different picture.
Both show similar performance, however they differ about 15.8\% in size.
Therefore, although a Java-specific pre-tokenization did not show a better performance, it does come with a reduced memory footprint.
Memory sensitive application may therefore profit from such approach.
Nonetheless, \emph{BERT Base Cased} shows a better performance with the same number of parameters as \emph{BERT Base Uncased} making it the obvious choice when considering performance.

With these observations, we can answer our research question stated in Section \ref{sec:ro}.
\paragraph*{RQ1} 
We have introduced an approach on data retrieval for certain programming languages, which in our case is Java.
The resulting performance of our trained models in comparison to a already pretrained model shows that our retrieval pipeline is able to deliver data for a state-of-the-art model.
Furthermore, as the pipeline is language-agnostic, we consider it also suitable for the code retrieval of other programming languages.
\paragraph*{RQ2}
We have investigated two different approaches to train a model upon software code.
An approach leveraging syntactical features of the Java language did not outperform the model using a vocabulary intended for natural language.
Therefore, we argue a tokenizer for NL performs well enough that a further optimization on this step may be neglected.
Comparing models trained solely for Java also shows a better performance than a model trained on several programming languages, showing the advantage of training solely on a single language.
Due to its performance, we choose the model using the \emph{BERT Base Cased} tokenizer as our final JavaBERT model.
\paragraph*{RQ3}
We employed a adapted WER metric to measure the performance of all trained models.
Our evaluation shows an accuracy of up to 94.4\% for prediction of masked tokens on the first try.
This accuracy value can be considered high enough to further use this model on SE related tasks.

\subsection{Threats to Validity}
The observed outcome of our experiment depends heavily on the used data and parameters.
Deviating from the used data may show different results in favor of other models.
We tried to circumvent this threat by using a broad set of data and by applying the same data to all models.
Also, we assume that GitHub provides data suitable for training tasks as used in our case.
The used hyperparameters are also a factor which impact the performance of each model.
Setting different values to these parameters may show different outcomes, which we avoided by using the same settings for all models, considering it to increase comparability.
Furthermore, we just evaluated a limited set of models, which may not be the best options for our use case.
Additionally, we assume the WER metric to be suitable for our evaluation and its scores to be valid indicator for the performance of each models.

\section{Future Work}
Future work based on our results can be divided into two categories.

First, our model may benefit from optimization approaches also used by the original BERT model.
\citet{Lan_Chen_Goodman_Gimpel_Sharma_Soricut_2020} optimized the original BERT model by applying techniques which effectively reduce the number of parameters within the model.
This reduces the memory footprint and accelerates the execution of the model.
Different model architectures may also be taken into consideration.
For instance, GPT2 and its causal language modeling introduce a different architecture and also a different generic training task, showing comparable performance\cite{radford2019language}.
In addition, a hyperparameter search was not conducted in this work, which may also optimize the performance of our model.

Second, we will focus on finetuning our model for more complex downstream tasks.
This may include the detection of malicious code, duplicate code detection, or comment generation.
Such tasks will require multimodal data retrieval approaches also showing the potential for future work upon the retrieval pipeline.
This will eventually allow us to introduce this model to the domain of SE.

\section{Conclusion}
Pretrained models for software code understanding are still scarce, although their uses in the domain of SE are manifold.
We therefore presented a concept on how to pretrain a model for a single programming language.
A data retrieval pipeline was used to showcase an approach on which syntactical features of programming languages are used within the pretraining process of multiple BERT models.
The same pipeline allowed us to train distinct models not relying on this technique and using a generic vocabulary instead.
While outperforming the language-specific models regarding accuracy on predictions for an MLM task, the latter vocabulary shows disadvantages in memory consumption and execution time.
We argue that the performance should be considered as the more relevant topic, making the model using a case-sensitive NL tokenizer the first choice for any tasks performed on Java software code.
All models pretrained on our collected data set outperform a model trained with a similar objective, showing state-of-the-art performance.
We therefore consider this model suitable for further research toward ML-based tooling for SE.
To encourage this, we uploaded the JavaBERT model to the Hugging Face Hub\footnote{\url{https://huggingface.co/CAUKiel/JavaBERT}} and the code for our retrieval pipeline to GitHub\footnote{\url{https://github.com/cau-se/gh-archive-code-retrieval}}.

\renewcommand*{\bibfont}{\small}

\end{document}